\documentclass[british,polish, 12pt]{article}
\usepackage[T1]{fontenc}
\usepackage[latin9]{inputenc}
\usepackage{color}
\usepackage{amsmath}
\usepackage{amssymb}
\usepackage{a4wide, amsmath, amsfonts, amssymb, mathrsfs, amsthm, bbm, color, stmaryrd, breqn}
\usepackage{esint}
\usepackage{amsfonts,color}

\makeatletter
\usepackage{babel}

\makeatother

 \global\long\def\sbr#1{\left[ #1\right] }
 \global\long\def\cbr#1{\left\{  #1\right\}  }
 \global\long\def\rbr#1{\left(#1\right)}
 
 \global\long\def\E{\mathbb{E}}
 \global\long\def\P{\mathbb{P}}
 \global\long\def\R{\mathbb{R}}
 \global\long\def\N{\mathbb{N}}

 \global\long\def\dd#1{\textnormal{d}#1}

 \global\long\def\ra{\rightarrow}

 \global\long\def\ns{\infty}

\newtheorem{theorem}{Theorem} \newtheorem{tw}[theorem]{Theorem}
\newtheorem{stw}[theorem]{Proposition} 
\newtheorem{rem}[theorem]{Remark} 
\newtheorem{definition}[theorem]{Definition}

\title{On SDEs with Lipschitz coefficients, driven by continuous, model-free
martingales}

\author{Lesiba Ch. Galane\thanks{University of Limpopo}, Rafa{\l } M.
\L ochowski\thanks{Warsaw School of Economics}  and Farai J. Mhlanga\thanks{University of Limpopo}}

\begin{document}

\selectlanguage{british}

\maketitle

\abstract{We prove the existence and uniqueness of solutions of SDEs with
Lipschitz coefficients, driven by continuous, model-free martingales.
The main tool in our reasoning is  Picard's iterative procedure and a model-free version of the Burkholder-Davis-Gundy
inequality for integrals driven by model-free, continuous martingales. We work with a new outer measure which assigns zero value exactly to those properties which are instantly blockable.}


\section{Introduction}

The main purpose of this paper is to prove the existence and uniqueness
of solutions of differential equations driven by continuous, model-free martingales. Continuous, model-free martingales were introduced in a recent book by Glenn Schafer and Vladimir Vovk \cite{ShaferVovk:2018}. Roughly speaking, model-free martingales are processes representing evolution of values of a dynamic portfolio consisting of several financial assets; they are related to model-free price paths (representing values of static portfolios consisting of one asset). \emph{Typical}, model-free price paths are processes representing evolution of prices of financial assets which do not allow to obtain infinite wealth during finite time by risking small amount and trading these assets.  From pioneering works of Vovk \cite{Vovk_volatility:2008},
\cite{Vovk_randomness:2009}, \cite{Vovk_probability:2012}, \cite{Vovk_cadlag:2015}
it is well known that typical model-free price paths reveal many
properties of local martingales. 
The case of continuous price paths is understood much better than
the case of c\`adl\`ag paths. 

However, even in the case of continuous, model-free price paths there are still
many topics which need to be understood better. One of such topics
is the existence and uniqueness of solutions of differential equations
driven by such paths. The first results in
this direction are proven in \cite{BartlKupperNeufeldSDE:2018}, even
for Hilbert space-valued processes. In \cite{BartlKupperNeufeldSDE:2018} the authors assume, similarly as we do here, that the coefficients of the differential equations are Lipschitz continuous, but they additionally assume some growth condition on the quadratic
variation process of the coordinate process, see \cite[Sect. 2, Remark 2.7]{BartlKupperNeufeldSDE:2018}. Another related paper is  \cite{LPP:2022}, where existence and uniqueness
result for one-dimensional differential equations, driven by typical paths, with non-Lipschitz
continuous coefficients in the spirit of Yamada-Watanabe as well as an approximation result
in the spirit of Doss-Sussmann were proven. 

Our approach is different. First, the driving processes of our equations are more general processes -- model-free, continuous martingales. Second, we work with the properties which hold with \emph{instant enforcement}. Roughly speaking, they are such properties that a trader (skeptic) is able to become infinitely rich as soon as they cease to hold, see \cite[Chapt. 14]{ShaferVovk:2018} and the next section. 

In this paper we will consider the following differential
equation (or rather integral equation)  driven by continuous, model-free, real martingales $X^1, X^2, \ldots X^d: \left[0,+\ns\right) \times \Omega \ra \R$:
\begin{equation}
{\mathbf Y}_{t}\left(\omega\right)={\mathbf Y}_{0}\left(\omega\right)+\int_{0}^{t}K\left(s,{\mathbf Y}\left(\omega\right),\omega\right)\dd {\mathbf A}_{s}+\int_{0}^{t}F\left(s,{\mathbf Y}\left(\omega\right),\omega\right)\dd{{\mathbf X}_{s}\left(\omega\right)},\label{eq:equation}
\end{equation}
where ${\mathbf A}:\left[0,+\ns\right)\times\Omega\ra\R^d$ is a continuous, adapted, finite-variation process, ${\mathbf X}$ is the vector-valued process with coordinates $X^1, X^2, \ldots X^d$, 
${\mathbf X} = \rbr{X^1, X^2, \ldots X^d}$, and $K, F:\left[0,+\ns\right)\times \rbr{\R^d}^{\left[0,+\ns\right)} \times\Omega\ra \R^{d \times d}$ are non-anticipating (the definition of non-anticipating functionals and formal statement of all assumptions
is given in Sect. \ref{sec:Picard's-iterations}), matrix-valued and  Lipschitz
in the sense that there exists $L\ge0$ such that for all $t\in\left[0,+\ns \right)$,
${\mathbf x}, {\mathbf y}:\left[0,+\ns \right)\ra\R^{d}$ and $\omega\in\Omega$
\begin{equation}
\left|K\left(t,{\mathbf x},\omega\right)-K\left(t,{\mathbf y},\omega\right)\right|+\left|F\left(t,{\mathbf x},\omega\right)-F\left(t,{\mathbf y},\omega\right)\right|\le L\sup_{s\in\sbr{0,t}}\left|{\mathbf x}\rbr s-{\mathbf y}\rbr s\right|,\label{eq:jedenF}
\end{equation}
where $\mid\cdot\mid$ denotes the Euclidean norm in $\R^n$ with $n = d\times d$ on the left side of \eqref{eq:jedenF} and $ n = d$ on the the right side of \eqref{eq:jedenF}, for example:
$
\left|K\left(t,{\mathbf x},\omega\right)-K\left(t,{\mathbf y},\omega\right)\right| = \rbr{\sum_{i,j = 1}^d \rbr{K^{i,j}\left(t,{\mathbf x},\omega\right)-K^{i,j}\left(t,{\mathbf y},\omega\right)}^2 }^{1/2}.
$

Equation \eqref{eq:equation} may be written as the system of integral equations: for $j =1,2,\ldots,  d$, 
 \begin{equation}
Y_{t}^j\left(\omega\right) = Y_{0}^j\left(\omega\right)+ \sum_{i=1}^d \int_{0}^{t}K^{i,j}\left(s,{\mathbf Y}\left(\omega\right),\omega\right)\dd {A}_{s}^i +\sum_{i=1}^d  \int_{0}^{t}F^{i,j}\left(s,{\mathbf Y}\left(\omega\right),\omega\right)\dd{{X}_{s}^i\left(\omega\right)} \label{eq:sys_equations}
\end{equation}
or, equivalently, 
 \begin{equation*}
Y_{t}^j\left(\omega\right) = Y_{0}^j\left(\omega\right)+ \int_{0}^{t}K^{j}\left(s,{\mathbf Y}\left(\omega\right),\omega\right)\dd {\mathbf A}_{s} +\int_{0}^{t}F^{j}\left(s,{\mathbf Y}\left(\omega\right),\omega\right)\dd{{\mathbf X}_{s}\left(\omega\right)},
\end{equation*}
where 
$\int_{0}^{t}K^{j}\left(s,{\mathbf Y}\left(\omega\right),\omega\right)\dd {\mathbf A}_{s}  = \sum_{i=1}^d \int_{0}^{t}K^{i,j}\left(s,{\mathbf Y}\left(\omega\right),\omega\right)\dd {A}_{s}^i$, $K^{j}\left(s,{\mathbf Y}\left(\omega\right),\omega\right) = \rbr{K^{i,j}\left(s,{\mathbf Y}\left(\omega\right),\omega\right)}_{i=1,2,\ldots, d}$ and a similar notation is used for $F$.
The integrals appearing in the first sum in (\ref{eq:sys_equations}) are understood as the standard Lebesgue-Stieltjes integrals, while integrals appearing in the second sum as model-free It\^o integrals introduced in the next section.

Condition (\ref{eq:jedenF}) is sufficient for our purpose. The same condition is used in \cite[Chapt. IX, Sect. 2]{RevuzYor:2005} but it differs from that used in \cite{BartlKupperNeufeldSDE:2018}.

This paper is organized as follows. In the next section we introduce
necessary definitions, notations and tools (like the model-free  BDG inequality). In the last
section we apply these tools and Picard's iterative procedure (used in a similar way as in \cite{DD1976}) to prove
the existence and uniqueness of the solution of (\ref{eq:equation}).

\section{Definitions, notation and auxiliary results}
First we outline a general setting in which we will work and which
follows closely \cite[Chapt. 14]{ShaferVovk:2018} and \cite{LochIE:2021}. $\N = \cbr{0,1,2, \ldots}$ is the set of positive integers and $b, d \in \N \setminus \cbr{0}$.  We will work with a martingale
space which is a quintuple
\[
\rbr{\Omega,{\cal F},\mathbb{F=}\rbr{{\cal F}_{t}}_{t\ge0},J=\cbr{1,2,\ldots,b},\cbr{S^{j},j\in J}}
\]
of the following objects: $\Omega$ is a space of possible outcomes
of reality, ${\cal F}$ is a $\sigma$-field of the subsets of $\Omega$
which we call \emph{events}, $\mathbb{F=}\rbr{{\cal F}_{t}}_{t\ge0}$
is a filtration and $\cbr{S^{j},j\in J}=\cbr{S^{1},S^{2},\ldots S^{b}}$
is a family of \emph{basic continuous martingales}, that is for any
$t\in[0,+\ns)$ and $j\in J$, $S_{t}^{j}$ is a $\rbr{{\cal F}_{t},{\cal B}(\R)}$-measurable
\emph{real variable} $S_{t}^{j}:\Omega\ra\R$ such that for each $\omega\in\Omega$
the trajectory $[0,+\ns)\ni t\mapsto S_{t}^{j}(\omega)$ is continuous
(${\cal B}(\R)$ denotes the $\sigma$-field of Borel subsets of $\R$).
Throughout the paper the filtration $\mathbb{F}$ is fixed, moreover,
we assume that ${\cal F}_{0}$ is trivial, ${\cal F}_{0}=\cbr{\emptyset,\Omega}$,
thus all $\rbr{{\cal F}_{0},{\cal B}(\R)}$-measurable variables $S_{0}^{j}$,
$j\in J$, are deterministic. 

A \emph{real process} $X:[0,+\ns)\times\Omega\ra\R$ is a collection
of real variables $X_{t}:\Omega\ra\R$, $t\in[0,+\ns)$, such that
$X_{t}$ is $\rbr{{\cal F}_{t},{\cal B}(\R)}$-measurable, thus all
processes which we consider are adapted to $\mathbb{F}$. 

A \emph{$d$-dimensional real process} $\mathbf Y$ is a $d$-tuple $\rbr{Y^1, Y^2, \ldots, Y^d }$ of real processes  $Y^1, Y^2, \ldots, Y^d$.

A \emph{process} $Y:[0,+\ns)\times\Omega\ra\R\cup\cbr{-\ns,+\ns}=[-\ns,+\ns]$,
is a collection of \emph{extended variables} $Y_{t}:\Omega\ra[-\ns,+\ns]$,
$t\in[0,+\ns)$, such that $Y_{t}$ is $\rbr{{\cal F}_{t},{\cal B}([-\ns,+\ns])}$-measurable
(any set in ${\cal B}([-\ns,+\ns])$ is of the form $A$, $A\cup\cbr{-\ns}$,
$A\cup\cbr{+\ns}$ or $A\cup\cbr{-\ns,+\ns}$, where $A\in{\cal B}(\R)$). 

A \emph{$d$-dimensional  process} $\mathbf Y$ is a $d$-tuple $\rbr{Y^1, Y^2, \ldots, Y^d }$ of processes  $Y^1, Y^2, \ldots, Y^d$.

A \emph{generalized process} is any function $Y:[0,+\ns)\times\Omega\ra[-\ns,+\ns]$.

For any generalized process $Y$ we define its supremum process $Y^*$, which
is a generalized process defined as
\[
Y_{t}^{*}(\omega):=\sup_{0\le s\le t}\left|Y_{t}(\omega)\right|,
\]
where we denote $Y_{t}(\omega):=Y(t,\omega)$. The generalized process $Y$ is \emph{globally bounded} iff $|Y_{t}(\omega)| < +\ns$ for all $(t, \omega) \in [0,+\ns)\times\Omega$.

Throughout the whole paper we apply the following convention. A sequence of real numbers $a_{n}$, where $n=0,1,2,\ldots$,
is denoted by $\rbr{a_{n}}$ or $\rbr{a_{n}}_{n}$ and a sequence
of real numbers $a^{n}$, where $n=0,1,2,\ldots$, is denoted by $\rbr{a^{n}}$
or $\rbr{a^{n}}_{n}$ (without indication that $n$ ranges over the
set of nonnegative integers $\N$). A similar convention will be applied
to infinite sequences of stopping times, variables etc.

A sequence of $\mathbb{F}$-stopping times $\rbr{\tau_{n}}$ is
called \emph{non-decreasing} if for all $n\in\N$ and \emph{each}
$\omega\in\Omega$, $\tau_{n+1}(\omega)\ge\tau_{n}(\omega)$. 

A sequence of $\mathbb{F}$-stopping times $\rbr{\tau_{n}}$ is
called \emph{proper} if it is non-decreasing, $\tau_{0}\equiv0$ and
for \emph{each} $\omega\in\Omega$ the sequence $\rbr{\tau_{n}(\omega)}$
is divergent to $+\ns$ or there exists some $n\in\N$ such that $\tau_{n}(\omega)=\tau_{n+1}(\omega)=\ldots\in[0,+\ns]$. 

A \emph{simple trading strategy} is a triplet $G=\rbr{c,\rbr{\tau_{n}},\rbr{g_{n}}}$
which consists of the initial capital $c\in\R$, a proper sequence
of $\mathbb{F}$-stopping times $\rbr{\tau_{n}}$ and a sequence
of $\rbr{{\cal F}_{\tau_{n}},{\cal B}\rbr{\R}}$-measurable real variables $g_{n}:\Omega\ra\R$, $n \in \N$, such
that $g_{n}(\omega)=0$ whenever $\tau_{n}(\omega)=+\ns$.

For a simple trading strategy $G=\rbr{c,\rbr{\tau_{n}},\rbr{g_{n}}}$  and a real process $X:[0,+\ns)\times\Omega\ra\R$ we define 
\begin{equation} \label{simple_int}
(G\cdot X)_{t}(\omega):=c+\sum_{n=1}^{+\ns}g_{n-1}(\omega)\rbr{X_{\tau_{n}(\omega)\wedge t}-X_{\tau_{n-1}(\omega)\wedge t}}.
\end{equation}
(For $s,t\in[-\ns,+\ns]$ we define $s\wedge t=\min\cbr{s,t}$.) Let us note that since the sequence $\rbr{\tau_n}$ is proper, there is only finite number of non-zero summands in the sum $\sum_{n=1}^{+\ns}g_{n-1}(\omega)\rbr{X_{\tau_{n}(\omega)\wedge t}-X_{\tau_{n-1}(\omega)\wedge t}}$ appearing in the definition of $(G\cdot X)_{t}(\omega)$.

We define the \emph{simple capital process} corresponding
to the vector ${\mathbf G} = \rbr{G^j}_{j \in J}$ of simple trading strategies $G^{j}$, $j\in J$, as 
\begin{equation} \label{simp_cap_process}
({\mathbf G} \cdot {\mathbf S})_{t}(\omega):=\sum_{j\in J}(G^{j}\cdot S^{j})_{t}(\omega).
\end{equation}
The simple capital process has a very natural interpretation -- it is the capital accumulated till time $t$ by the application
of the simple trading strategy $G^j$ to the asset whose price is equal to the basic martingale $S^j$, $j \in J$.

The class ${\cal C}$ of \emph{nonnegative supermartingales} is defined
as the smallest $\liminf$-closed class of processes containig all simple capital
processes which are non-negative, that is ${\cal C}$ contains all
nonnegative simple capital processes and for any sequence $\rbr{X^{n}}$
such that $X^{n}\in{\cal C}$ for $n\in\N$, we have that $X:=\liminf_{n\ra+\ns}X^{n}$
also belongs to ${\cal C}$. 

A property $E\subseteq[0,+\ns)\times\Omega$
is \emph{instantly enforceable}, or holds \emph{with instant enforcement}, w.i.e. in short, if there exists a nonnegative supermartingale
$X$ such that $X_{0}=1$ and 
\[
(t,\omega)\notin E\Longrightarrow X_{t}(\omega)=+\ns.
\]
Complements of instantly enforceable properties (sets) are called
\emph{instantly blockable}. 

We define \emph{upper expectation} (or cost of super-hedging
or super-replication) of a generalized process $Y:[0,+\ns)\times\Omega\ra[-\ns,+\ns]$
in the following way
\begin{align*}
\overline{\E}Y := \inf & \left\{ \lambda\in\R:\exists X\in{\cal C}\text{ such that }\forall(t,\omega)\in[0,+\ns)\times\Omega,  X_{0}(\omega)\le\lambda \right. \\ & \left. \text{ and }X_{t}(\omega)\ge Y_{t}(\omega)\right\}
\end{align*}
and for $A\subseteq[0,+\ns)\times\Omega$ we define its \emph{outer measure} as
$
\overline{\P}(A)=\overline{\E}{\mathbf 1}_{A}.
$
We have the following result (see \cite[Lemma 2.1]{LochIE:2021}).
\begin{stw} The set $B\subseteq[0,+\ns)\times\Omega$ is instantly
blockable iff
$
\overline{\P}(B)=0.
$
\end{stw}

Next to the class of nonnegative supermartingales, other important
class of processes which we will work with is the family of \emph{martingales}.
The class of martingales ${\cal M}$ is defined as the smallest $\lim$-closed
class of real process such that it contains all simple capital processes.
By the fact that ${\cal M}$ is $\lim$-closed we mean that whenever
$X^{n}\in{\cal M}$, $n\in\N$, and  $X$ is a real process such that  for any $(t,\omega)\in[0,+\ns)\times\Omega$, 
\begin{equation}
\lim_{n\ra+\ns}\sup_{s\in[0,t]}\left|X_{s}(\omega)-X_{s}^{n}(\omega)\right|=0\text{ w.i.e.}\label{eq:unifconv}
\end{equation}
then also $X\in{\cal M}$. Condition (\ref{eq:unifconv}) guarantees
that the limit process $X$ is continuous w.i.e.

For any $X \in {\cal M}$ there exists its \emph{quadratic variation} process denoted by $\sbr{X}$ (see  \cite[Proposition 4.3]{LochIE:2021}), which is non-decreasing and continuous w.i.e. (and one may take a version which is non-decreasing and continuous for all $\omega \in \Omega$), and for any $p\ge 1$ the following BDG inequalities hold (see  \cite[Proposition 4.6]{LochIE:2021}):
\[
c_{p}\overline{\E}\sbr X^{p/2}\le\overline{\E}\rbr{\rbr{X-X_{0}}^{*}}^{p}\le C_{p}\overline{\E}\sbr X^{p/2}.
\]
In the case $p>1$ one may take $C_{p}=6^{p}(p-1)^{p-1}$ and $c_{p}=1/C_{p}$,
while in the case $p=1$ one may take $C_{p}=6$ and $c_{p}=1/3$.

Until now we have not defined integrals appearing in \eqref{eq:sys_equations}. 
In  \cite{{ShaferVovk:2018}} and \cite{LochIE:2021} there were defined integrals with respect to model-free, continuous martingales, but now we will define integrals suiting our needs better. 
To do this let us introduce  the spaces $\mathcal G^0 = \mathcal G_{\mathbf X}^0$ and $\mathcal H = \mathcal H_{\mathbf X}$ of (equivalence classes of)  $d$-dimensional processes and processes respectively, equipped with the norms:
\[
\|\mathbf Y \|_{\ns,{\mathbf X},loc}^{\mathcal G} := \sum_{N=1}^{\ns} 2^{-N} \overline{\E} |\mathbf Y|_{\cdot \wedge \sigma({\mathbf X},N)}^*, \ \| Z \|_{\ns,{\mathbf X},loc}^{\mathcal H} := \sum_{N=1}^{\ns} 2^{-2N} \overline{\E} |Z|_{\cdot \wedge \sigma({\mathbf X},N)}^*
\]
where ${\mathbf Y} : [0, +\ns) \times \Omega \ra [-\ns,+\ns]^d$, $|Y| = \sqrt{\sum_{i=1}^d \rbr{Y^j}^2}$ and we define 
\[
\sigma(\mathbf X,N) := \sigma\rbr{\sbr{X^1},N} \wedge \sigma\rbr{\sbr{X^2},N} \wedge \ldots \wedge \sigma\rbr{\sbr{X^d},N},
\]
where for the martingale $X$, $$\sigma\rbr{\sbr{X},N} = \inf\cbr{t \ge 0: \sbr{X}_t \ge N}.$$
To deal with the values $+\ns$ and $-\ns$ which may be attained by some components of (a representative of) some element $\mathbf Y$ of $\mathcal G$ we apply the convention that $+\ns - (+\ns) = 0$ and  $-\ns - (-\ns) = 0$. 
Further, let $\mathcal G = \mathcal G_{\mathbf X}$ be a closure of the linear subspace of $\mathcal G^0$ spanned by c\`adl\`ag \emph{$d$-dimensional step processes} of the form ${\mathbf G} = \rbr{G^i}_{i=1,2,\ldots, d}$, where
\[
G_t^i(\omega) := \sum_{n=1}^{+\ns}g_{n-1}^i(\omega){\mathbf 1}_{\left[\tau_{n-1}^i(\omega),  \tau_{n}^i(\omega)\right)}(t),
\]
and $G^i = \rbr{0,\rbr{\tau_{n}^i},\rbr{g_{n}^i}}$, $i=1,2,\ldots, d$, are {simple trading strategies}. For the c\`adl\`ag $d$-dimensional step process ${\mathbf G}$ we define the \emph{simple integral process} as 
$$\mathbf G\cdot \mathbf X =  \sum_{i=1}^d G^i \cdot X^i,$$ where 
$G^i \cdot X^i $ are defined by \eqref{simple_int}.
Instead of $G^i \cdot X^i $ we will also write $\int_0^{\cdot} G_s^i \dd X_s^i$ and instead of $\rbr{G^i \cdot X^i }_t$ we will also write 
$\int_0^{t} G_s^i \dd X_s^i$.
\begin{rem}
{More appropriate notation (consistent with the Stieltjes integral) to denote simple capital processes defined in \eqref{simp_cap_process} and just defined simple integrals, would be ${\mathbf G}_- \cdot {\mathbf S}$ and ${\mathbf G}_- \cdot {\mathbf X}$ respectively. Similarly, more appropriate notation in \eqref{eq:equation} would be $\int_{(0,t]}K\left(s-,{\mathbf Y}\left(\omega\right),\omega\right)\dd {\mathbf A}_{s}$ and $\int_{(0,t]}F\left(s-,{\mathbf Y}\left(\omega\right),\omega\right)\dd{{\mathbf X}_{s}\left(\omega\right)}$, but we will not use it to be consistent with the notation used in \cite{{ShaferVovk:2018}} and \cite{LochIE:2021}.}
\end{rem}
In the sequel we will  also use the fact that for a simple trading strategy $G$,  
$
\sbr{G \cdot X } = \int_0^{\cdot} \rbr{G_s}^2 \dd \sbr{X}_s
$ w.i.e., see \cite[Fact 5.1]{LochIE:2021}.
\begin{stw} \label{int_definition}
The spaces $\mathcal G$ and  $\mathcal H$ are Banach spaces. Two processes $\mathbf Y^1$  and $\mathbf Y^2$ are representatives of the same classes in $\mathcal G$ iff  $\mathbf Y^1= \mathbf Y^2 $ w.i.e., which is equivalent with $\overline{\E} |\mathbf Y^1-\mathbf Y^2| = 0$. A similar statement holds for processes in $\mathcal H$.
For any $\mathbf G \in \mathcal G$ which is the limit of $d$-dimensional step processes $\mathbf G^n$ in $ \mathcal G$, there exists the limit of $\mathbf G^n \cdot \mathbf X$ in $\mathcal H$ and we define $\mathbf G \cdot \mathbf X$ as this limit. Moreover, $\mathbf G \cdot \mathbf X$ has a representative in $\mathcal H$ which is a martingale, which implies that any representative of $\mathbf G \cdot \mathbf X$ in $\mathcal H$ is a martingale.
\end{stw}
\begin{proof}

The proof that $\|\mathbf \cdot \|_{\ns,{\mathbf X},loc}^{\mathcal G}$ defines a metric and that two processes $\mathbf Y^1$  and $\mathbf Y^2$ are representatives of the same classes in $\mathcal G$ iff  $\mathbf Y^1= \mathbf Y^2 $ w.i.e., which is equivalent with $\overline{\E} |\mathbf Y^1-\mathbf Y^2| = 0$, is omitted. To prove the completeness let $\rbr{Y^n}$ be a Cauchy sequence with respect to the metric $d_{\ns,\mathbf X,loc}^{\mathcal G}$ induced by the norm $\|\mathbf \cdot \|_{\ns,{\mathbf X},loc}^{\mathcal G}$. Let $\rbr{d_k}$ be any sequence of positive reals such that  $\sum_{k=1}^{+\ns} d_k  < +\ns$. There exists a subsequence $\rbr{\mathbf Y^{n_k}}$ such that for  $n \ge n_k$, $n,k =1,2,\ldots$ one has $d_{\ns,\mathbf X,loc}^{\mathcal G}\rbr{\mathbf Y^{n},\mathbf Y^{n_{k}}} \le d_k$. Taking $\mathbf Y := \liminf_{l \ra + \ns} \mathbf Y^{n_l}$, for $n \ge n_k$ we get 
\[
d_{\ns,\mathbf X,loc}^{\mathcal G}\rbr{\mathbf Y^{n}, \mathbf Y} \le d_{\ns,\mathbf X,loc}^{\mathcal G}\rbr{\mathbf Y^{n}, \mathbf Y^{n_k}} +  \sum_{l=k}^{+\ns} d_{\ns,\mathbf X,loc}^{\mathcal G}\rbr{\mathbf Y^{n_l},\mathbf Y^{n_{l+1}}} \le d_k + \sum_{l=k}^{+\ns} d_l,
\]
thus $\mathbf Y$ is the limit of the sequence $\rbr{\mathbf Y^n}$ (as a limit one can also take $\limsup_{l \ra + \ns} \mathbf Y^{n_l}$).
Similarly we prove the completeness of $\mathcal H$.

For two step processes $\mathbf G^m$  and $\mathbf G^n$, using the BDG inequalities, we estimate 
\begin{align*}
&\overline{\E} |\mathbf G^m\cdot \mathbf X - \mathbf G^n\cdot \mathbf X|_{\cdot \wedge \sigma({\mathbf X},N)}^* \le \sum_{i=1}^d \overline{\E} | G^{m,i}\cdot X^i - G^{n,i}\cdot X^i|_{\cdot \wedge \sigma({\mathbf X},N)}^* \\
& \le C_1 \sum_{i=1}^d \overline{\E} \rbr{\int_0^{\cdot \wedge \sigma({\mathbf X},N)}\rbr{G^{m,i}-G^{n,i}}^2_s \dd \sbr{X^i}_s}^{1/2} \\
& \le C_1 \sum_{i=1}^d \overline{\E} \rbr{\rbr{G^{m,i}-G^{n,i}}_{\cdot \wedge\sigma({\mathbf X},N)}^{*}  N^{1/2}}  \\ 
& = C_1  N^{1/2} \sum_{i=1}^d \overline{\E} \rbr{G^{m,i}-G^{n,i}}_{\cdot \wedge\sigma({\mathbf X},N)}^{*} \le C_1  N^{1/2} d \cdot \overline{\E} \rbr{ \mathbf G^m - \mathbf G^n}_{\cdot \wedge\sigma({\mathbf X},N)}^{*} \\ 
& \le C_1  N^{1/2} d  2^N \| \mathbf G^m - \mathbf G^n \|_{\ns,{\mathbf X},loc}^{\mathcal G}  .
\end{align*}
From the last estimate it follows that $\rbr{\mathbf G^n\cdot \mathbf X}$ is a Cauchy sequence in $\mathcal H$, since
\begin{align*}
&\| \mathbf G^m\cdot \mathbf X - \mathbf G^n\cdot \mathbf X \|_{\ns,{\mathbf X},loc}^{\mathcal H} := \sum_{N=1}^{\ns} 2^{-2N} \overline{\E} |\mathbf G^m\cdot \mathbf X - \mathbf G^n\cdot \mathbf X|_{\cdot \wedge \sigma({\mathbf X},N)}^* \\
&\le \sum_{N=1}^{\ns} 2^{-2N}  C_1  N^{1/2} d  2^N \|\mathbf G^m - \mathbf G^n \|_{\ns,{\mathbf X},loc}^{\mathcal G}  
= \rbr{C_1 d \sum_{N=1}^{\ns} 2^{-N}   N^{1/2} } \|\mathbf G^m - \mathbf G^n \|_{\ns,{\mathbf X},loc}^{\mathcal G}.
\end{align*}
$\mathbf G \cdot \mathbf X$ is a limit in $\mathcal H$ of $\mathbf G^n \cdot \mathbf X$, which are martingales. To prove that it has a representative in $\mathcal H$ which is a martingale let $\left(n_k\right)_k$ be any subsequence of the sequence of all natural numbers such that $ M := \sum_{k=1}^{+\infty} \| \mathbf G \cdot \mathbf X - \mathbf G^{n_k}\cdot \mathbf X \|_{\infty,{\mathbf X},loc}^{\mathcal H}  < +\infty$ and let  $B\subseteq [0,+\infty)\times\Omega$ be the set of $(t, \omega)$ where  $\left(\mathbf G \cdot \mathbf X - \mathbf G^{n_k} \cdot \mathbf X \right)_t^*(\omega) \nrightarrow 0$. Let $(t, \omega) \in B$ and $N \in \mathbb N$ be such that $\sigma(\mathbf X,N)(\omega) \ge t $. We have 
\[
\sum_{k=1}^{+\infty}  \left(\mathbf G \cdot \mathbf X - \mathbf G^{n_k} \cdot \mathbf X \right)_{\sigma(\mathbf X,N)(\omega)}^*(\omega)  \ge  \sum_{k=1}^{+\infty}  \left(\mathbf G \cdot \mathbf X - \mathbf G^{n_k} \cdot \mathbf X \right)_t^*(\omega) = +\infty.
\]
As a result, for any $\varepsilon >0$
\[
\varepsilon \sum_{k=1}^{+\infty} \sum_{N=1}^{+\infty} 2^{-2N} \left(\mathbf G \cdot \mathbf X - \mathbf G^{n_k} \cdot \mathbf X \right)_{\sigma(\mathbf X,N)(\omega)}^*(\omega) = +\infty.
\]
On the other hand, since 
\begin{align*}
& \overline{\mathbb E} \sum_{k=1}^{+\infty} \sum_{N=1}^{+\infty} 2^{-2N} \left(\mathbf G \cdot \mathbf X - \mathbf G^{n_k} \cdot \mathbf X \right)_{\cdot \wedge \sigma(\mathbf X,N)}^* \le  \sum_{k=1}^{+\infty}  \overline{\mathbb E} \sum_{N=1}^{+\infty} 2^{-2N} \left(\mathbf G \cdot \mathbf X - \mathbf G^{n_k} \cdot \mathbf X \right)_{\cdot \wedge \sigma(\mathbf X,N)}^* \\
& = \sum_{k=1}^{+\infty}  \| \mathbf G \cdot \mathbf X - \mathbf G^{n_k}\cdot \mathbf X \|_{\infty,{\mathbf X},loc}^{\mathcal H} = M  < +\infty,
\end{align*}
we know that there exist a non-negative supermartingale which starts from a capital no greater than $\varepsilon M$ and attains value $+\infty$ on $B$. Since $\varepsilon$ is arbitrary positive real, $B$ is  instantly blockable, which implies that $\mathbf G \cdot \mathbf X$ is a martingale. 
\end{proof}

\section{Theorem on existence and uniqueness of the solutions of SDEs with
Lipschitz coefficients, driven by continuous, model-free martingales
\label{sec:Picard's-iterations}}

In this section we prove the existence and uniqueness of the solution
of SDE (\ref{eq:equation}). We will assume the following:
\begin{enumerate}
\item { $A^{j}=A^{j,u}-A^{j,v}$, $j = 1,2, \ldots, d$,
and $|A|^{j} = A^{j,u} + A^{j,v}$, $|A| = \rbr{\sum_{j=1}^d \rbr{|A|^{j}}^2}^{1/2}$, where 
$A^{j,u},A^{j,v}:\left[0,+\ns\right)\times\Omega\ra\R$ are continuous, non-decreasing, adapted processes, starting from $0$};
\item $K, F:\left[0,+\ns\right)\times \rbr{\R^d}^{\left[0,+\ns\right)}\times\Omega\ra\R^{d\times d}$, and $K$ 
and $F$ are \emph{non-anticipating}, by which we mean that 
\begin{enumerate} 
\item for any $t\in\left[0,+\ns\right)$, $\omega \in \Omega$ and two functions ${\mathbf x}, {\mathbf y}:\left[0,+\ns\right)\ra\R^{d}$,
$K\rbr{t,{\mathbf x},\omega}=K\rbr{t,{\mathbf y},\omega}$ and
$F\rbr{t,{\mathbf x},\omega}=F\rbr{t,{\mathbf y},\omega}$ whenever
${\mathbf x}(s)={\mathbf y}(s)$ for all $s\in\sbr{0,t}$; 
\item {for any adapted c\`adl\`ag process ${\mathbf Y}:\left[0,+\ns\right)\times\Omega \ra \R^d$ the processes $K_t (\omega) := K\rbr{t,{\mathbf Y}(\omega),\omega}$, $F_t (\omega) := F\rbr{t,{\mathbf Y}(\omega),\omega}$ are adapted and c\`adl\`ag};
\end{enumerate}
\item $K$ and $F$ satisfy condition (\ref{eq:jedenF}).
\end{enumerate}
For a  process ${\mathbf Y}:\left[0,+\ns\right)\times\Omega \ra \R^d$ instead of $K\left(s,{\mathbf Y}\left(\omega\right),\omega\right)$ we will often write $K\left(s,{\mathbf Y}\right)$ and instead of 
$F\left(s,{\mathbf Y}\left(\omega\right),\omega\right)$
we will often write $F\left(s,{\mathbf Y}\right)$.

Now let us define what we will mean by the solution of \eqref{eq:equation}. 
\begin{definition} \label{solution_def}
A solution of \eqref{eq:equation} is any $d$-dimensional process $\mathbf Y$ such that there exist a sequence of non-decreasing $\mathbb{F}$-stopping times $\rbr{\tau_{n}}$, which tend to $+\ns$ for all $\omega \in \Omega$ and such that $\mathbf Y_{\cdot \wedge \tau_n}$ ($\mathbf Y$ stopped at the time $\tau_n$) is a representative of some element of $\mathcal G$ and the following equalities for $j=1,2,\ldots,d$, $n \in \N$ and  any $t \in [0, +\ns)$ hold:
\begin{align} 
 Y_{t \wedge \tau_n}^j - Y_{0}^j -  \int_{0}^{t}K^{j}\left(s,{\mathbf Y_{t \wedge \tau_n}}\right) {\mathbf 1}_{[0, \tau_n)}(s) \dd {\mathbf A}_{s} = \int_{0}^{t}F^{j}\left(s,{\mathbf Y_{t \wedge \tau_n}}\right){\mathbf 1}_{[0, \tau_n)}(s) \dd {\mathbf X}_{s}, \label{meaning}
\end{align}
where the integral on the left hand side of \eqref{meaning} is understood as the usual Lebesque-Stieltjes integral while the  integral on the right hand side of \eqref{meaning}  as the integral defined in Proposition \ref{int_definition} (by the Lipschitz assumption \eqref{eq:jedenF},  the $d$-dimensional processes $F^j\left(\cdot,{\mathbf Y_{\cdot \wedge \tau_n}}\right)$, $j=1,2,\ldots, d$, belong to (are representatives of some elements of) $\mathcal G$ and the integrals $\int_{0}^{t}F^{j}\left(s,{\mathbf Y_{t \wedge \tau_n}}\right){\mathbf 1}_{[0, \tau_n)}(s) \dd{{\mathbf X}_{s}}$ are well defined). The equality in \eqref{meaning} is understood as the fact that the process $Y_{t \wedge \tau_n}^j - Y_{0}^j -  \int_{0}^{t}K^{j}\left(s,{\mathbf Y_{t \wedge \tau_n}}\right) {\mathbf 1}_{[0, \tau_n)}(s) \dd {\mathbf A}_{s} $ is a representative  of the same  equivalence class in $\mathcal H$ which is on the  right hand side of \eqref{meaning}.
\end{definition}

We will use a model-free version of the BDG inequality for $p=1$ and Picard's iterative procedure (used in a similar way as in \cite{DD1976}) to prove the following theorem.

\begin{tw} \label{theorem_SDE} Under the assumptions 1.-3. stated above,
integral equation (\ref{eq:equation}) has a solution in the sense of Definition \ref{solution_def} and this solution is unique in the sense that for any two solutions $\mathbf G$ and $\mathbf H$ we have $\overline{\E}  \rbr{\mathbf G - \mathbf H}^* = 0$, or, eqiuvalently, $\mathbf G = \mathbf H$ w.i.e.
\end{tw}
\begin{rem}
Theorem \ref{theorem_SDE} implies the existence of a solution of (\ref{eq:equation}) in the sense of Definition \ref{solution_def}.
Naturally, for many equations, like for example the one-dimensional Black-Scholes equation
$ Y_t = y_0 + \int_0^t Y_s \dd A_s + \sigma \int_0^t Y_s \dd X_s
$
($x_0, \sigma$ - deterministic) we can write the solution explicitly
$
Y_t = y_0 \exp\rbr{ {A_t -\frac{1}{2}\sigma^2\sbr{X}_t} +  \sigma \rbr{X_t-X_0}}
$
and verify that it satisfies the Black-Scholes equation using the (model-free) It\^o formula (see \cite{Vovk_cadlag:2015}). However, for more general equations we often have no explicit solutions and the existence of a solution is not obvious.
\end{rem}

\subsection{Proof of Theorem \ref{theorem_SDE}}

\subsubsection{Existence}

Let us define $q = {1}/\rbr{3L}$, $r =  {1}/\rbr{3C_1d^2 L}$,
\[
\vartheta_{0}:=\inf\{t\ge0 : |A|_t \geq q\}, \ \sigma_{0}:=\inf\left\{ t\ge0 :\max_{j=1,2,\ldots,d} \sbr{X^j}_{t} \ge r \right\},
\]
\[
\theta_{0}=\vartheta_{0}\wedge\sigma_{0}
\]
and for any $\mathbf G$ in $\mathcal G$ define 
\begin{align*}
\left(T^{0}{\mathbf G}\right)_{t} & ={\mathbf Y}_{0}+\int_{0}^{t\wedge\theta_{0}}K\left(s, {\mathbf G}\right)\dd {\mathbf A}_{s}+\int_{0}^{t\wedge\theta_{0}}F\left(s, {\mathbf G}\right)\dd{{\mathbf X}_{s}} \\
& = {\mathbf Y}_{0}+\int_{0}^{t}K\left(s, {\mathbf G}\right) \mathbf 1_{\left[0, \theta_{0}\right)}(s)\dd {\mathbf A}_{s}+\int_{0}^{t}F\left(s, {\mathbf G}\right)1_{\left[0, \theta_{0}\right)}(s) \dd{{\mathbf X}_{s}}.
\end{align*}
Let us fix $N \in \N$. By the inequality 
\begin{align*}
& \rbr{ \int_{0}^{\cdot \wedge\theta_{0} \wedge \sigma({\mathbf X},N)} F\left(s, {\mathbf G}\right)\dd{{\mathbf X}_{s}} }^* \\
 & \le \sum_{i=1}^d \left(\sum_{j=1}^d  \int_{0}^{\cdot \wedge\theta_{0} \wedge \sigma({\mathbf X},N)}\left\{ F^{i,j}\left(s,{\mathbf G}\right)-F^{i,j}\left(s,{\mathbf H}\right)\right\} \dd{X_{s}^j}\right)^{*} \\
& \le \sum_{i,j=1}^d  \left(\int_{0}^{\cdot \wedge\theta_{0} \wedge \sigma({\mathbf X},N)}\left\{ F^{i,j}\left(s,{\mathbf G}\right)-F^{i,j}\left(s,{\mathbf H}\right)\right\} \dd{X_{s}^j}\right)^{*}
\end{align*}
(which follows from the estimate $\sqrt{\sum_{i=1}^d a_i^2} \le \sum_{i=1}^d |a_i|$), 
the subadditivity of $\overline{\E}$, the Lipschitz property and the BDG inequality, for any $\mathbf G, \mathbf H \in \mathcal G$ we estimate
\begin{align}
& \overline{\E}\left(T^{0}{\mathbf G}-T^{0}{\mathbf H}\right)^{*}_{\cdot \wedge \sigma({\mathbf X},N)} \nonumber \\
 & \le \overline{\E}\left(\int_{0}^{\cdot \wedge\theta_{0} \wedge \sigma({\mathbf X},N)} | K\left(s,{\mathbf G}\right)-K \left(s,{\mathbf H}\right)| \ \dd |A|_{s}\right)^{*}\nonumber \\
 & \quad+\sum_{i,j=1}^d  \overline{\E}\left( \int_{0}^{\cdot \wedge\theta_{0} \wedge \sigma({\mathbf X},N)}\left\{ F^{i,j}\left(s,{\mathbf G}\right)-F^{i,j}\left(s,{\mathbf H}\right)\right\} \dd{X_{s}^j}\right)^{*}\nonumber \\
   &  \le \overline{\E}\left(\int_{0}^{\cdot \wedge\theta_{0} \wedge \sigma({\mathbf X},N)} L \rbr{{\mathbf G}-{\mathbf H}}_s^* \ \dd |A|_{s}\right)^{*}\nonumber \\
 & \quad+C_1 \sum_{i,j=1}^d  \overline{\E}\left( \int_{0}^{\cdot \wedge\theta_{0} \wedge \sigma({\mathbf X},N)} L^2 \rbr{\rbr{{\mathbf G}-{\mathbf H}}_s^*}^2 \dd \sbr{X^j}_{s} \right)^{1/2} \nonumber \\
   &  \le \overline{\E} \left( L \rbr{{\mathbf G}-{\mathbf H}}^*_{\cdot \wedge \sigma({\mathbf X},N)} |A|_{\cdot \wedge\theta_{0}} \right) \nonumber \\
 & \quad+ C_1 \sum_{i,j=1}^d  \overline{\E}  \left( L  \rbr{{\mathbf G}-{\mathbf H}}^*_{\cdot \wedge \sigma({\mathbf X},N)} \rbr{ \sbr{X^j}_{\cdot \wedge\theta_{0}}}^{1/2} \right) \nonumber \\
 &  \le \overline{\E} \left( L \rbr{{\mathbf G}-{\mathbf H}}^*_{\cdot \wedge \sigma({\mathbf X},N)} \frac{1}{3L} \right) 
 + C_1 \sum_{i,j=1}^d  \overline{\E}  \left( L  \rbr{{\mathbf G}-{\mathbf H}}^*_{\cdot \wedge \sigma({\mathbf X},N)} \frac{1}{3C_1d^2 L}  \right) \nonumber \\
 &  = \frac{2}{3} \overline{\E} \rbr{{\mathbf G}-{\mathbf H}}^*_{\cdot \wedge \sigma({\mathbf X},N)}  \label{eq:Lipschitz}.
\end{align}
We have also used the fact that for $\mathbf G \in \mathcal G$, $i,j = 1,2,\ldots, d$, $G^i \cdot X^j $ is a martingale (which follows from Proposition \ref{int_definition}) and that 
$\left[G^i \cdot X^j \right] = \int_0^{\cdot} \left(G_s^i\right)^2 \,\mathrm{d} [X^j]_s
$ w.i.e., which follows from \cite[Fact 5.1]{LochIE:2021} by simple approximation arguments (see also \cite[Fact 5.5]{LochIE:2021}).

From \eqref{eq:Lipschitz} we get 
\begin{equation} \label{eq:Lipschitz1}
\| T^{0}\mathbf G - T^{0}\mathbf H \|_{\ns, {\mathbf X}, loc}^{\mathcal G} \le \frac{2}{3} \| \mathbf G - \mathbf H \|_{\ns, {\mathbf X}, loc}^{\mathcal G}.
\end{equation}
By \eqref{eq:Lipschitz1}, $\| T^{0}\mathbf G \|_{\ns, {\mathbf X}, loc}^{\mathcal G} \le \frac{2}{3} \| \mathbf G \|_{\ns, {\mathbf X}, loc}^{\mathcal G} < +\ns.$ We will now show that $T^{0}\mathbf G$ is a limit of step processes in $\mathcal G$ from which it will follow that $T^{0}\mathbf G \in \mathcal G$ ($T^{0}\mathbf G$ is a representative of an element of $\mathcal G$).
First, notice that $\mathbf G \in \mathcal G$ thus it is a limit (in $\mathcal G$) of a sequence of $d$-dimensional step processes $\mathbf G^n \in \mathcal G$, $n \in \N$. By \eqref{eq:Lipschitz1}, $$\| T^{0}\mathbf G^n \|_{\ns, {\mathbf X}, loc}^{\mathcal G} \le \frac{2}{3} \| \mathbf G^n \|_{\ns, {\mathbf X}, loc}^{\mathcal G} \le \| \mathbf G \|_{\ns, {\mathbf X}, loc}^{\mathcal G}$$ for sufficiently large $n$.  For each step (thus c\`adl\`ag) process $\mathbf G^n$, $K\left(s,{\mathbf G^n}\right)$ and $F\left(s,{\mathbf G^n}\right)$  are again adapted c\`adl\`ag processes (assumption 2(b)) which may be uniformly approximated by step processes $K^n$ and $F^n$ with given accuracy $\varepsilon >0$, respectively. 
For example if we define 
$\tau_0^{n, \varepsilon} := 0$,  $f_0^{n, \varepsilon} := 0$ and for $m \in \N\setminus \cbr{0} $ 
\[
\tau_m^{n, \varepsilon} := \inf \cbr{t > \tau_{m-1}^{n, \varepsilon} : |F^n_t - f_{m-1}^{n, \varepsilon}| \ge  \varepsilon }, \quad f_{m}^{n, \varepsilon} = F^n_{\tau_{m}^{n, \varepsilon}}, 
\]
then  $F^{n, \varepsilon}_t := \sum_{m=1}^{+\ns} f^{n, \varepsilon}_{m-1} {\mathbf 1}_{\left[\tau^{n, \varepsilon}_{m-1}, \tau^{n, \varepsilon}_m \right)} (t)$ approximates $F^n$ uniformly on $[0,+\ns)$ with accuracy $\varepsilon$, that is 
$
\sup_{t \in [0,+\ns)} |F^{n, \varepsilon}_t  - F^n_t| \le \varepsilon.
$ 
The integrals $\int_{0}^{\cdot \wedge\theta_{0}}K^n \dd {\mathbf A}_{s}$ and $\int_{0}^{\cdot \wedge\theta_{0}}F^n \dd {\mathbf X}_{s}$ are continuous. We also have the estimate 
\begin{align*}
& \overline{\E}\left(\int_{0}^{\cdot \wedge\theta_{0}}K\left(s,{\mathbf G^n}\right)  \dd {\mathbf A}_{s}  - \int_{0}^{\cdot \wedge\theta_{0}}K^n \dd {\mathbf A}_{s} \right)^{*}_{\cdot \wedge \sigma({\mathbf X},N)} \\
& \le \overline{\E}\left(\int_{0}^{\cdot \wedge\theta_{0} \wedge \sigma({\mathbf X},N)} \varepsilon \ \dd |A|_{s}\right)^{*}\nonumber \le \varepsilon q.
\end{align*}
and by the BDG inequality we estimate 
\begin{align*}
& \overline{\E}\left(\int_{0}^{\cdot \wedge\theta_{0}}F\left(s,{\mathbf G^n}\right)  \dd {\mathbf X}_{s}  - \int_{0}^{\cdot \wedge\theta_{0}}F^n \dd {\mathbf X}_{s} \right)^{*}_{\cdot \wedge \sigma({\mathbf X},N)} \\
& \le C_1 \sum_{i,j=1}^d  \overline{\E} \left( \int_{0}^{\cdot \wedge\theta_{0} \wedge \sigma({\mathbf X},N)} \varepsilon^2 \dd \sbr{X^j}_{s} \right)^{1/2}  \le C_1 d^2 \varepsilon r.
\end{align*}
From last two inequalities we infer that  $\|T^{0}\mathbf G^n - \mathbf Y_0 -  \int_{0}^{\cdot \wedge\theta_{0}}K^n \dd {\mathbf A}_{s}  - \int_{0}^{\cdot \wedge\theta_{0}}F^n \dd {\mathbf X}_{s} \|_{\ns, {\mathbf X}, loc}^{\mathcal G}$ may be as small as we please, thus  $T^{0}\mathbf G^n$ may be approximated with arbitrary accuracy by continuous processes in $\mathcal G$, thus the same holds for $T^{0}\mathbf G$, thus $T^{0}\mathbf G \in \mathcal G$.

Now we know that $T^0$ may be viewed as a mapping $T^0 :  \mathcal G \ra  \mathcal G$, which by \eqref{eq:Lipschitz1} is a contraction.
This contraction has a unique fixed point $\mathbf Y^{0}$ which for any $t \in [0,  +\ns) $ satisfies 
\begin{align*}
{\mathbf Y}^0_{t \wedge \theta_0} & ={\mathbf Y}_{0}+\int_{0}^{t\wedge \theta_0}K\left(s, {\mathbf Y^0}\right)\dd {\mathbf A}_{s}+\int_{0}^{t\wedge \theta_0}F\left(s, {\mathbf Y^0}\right)\dd{{\mathbf X}_{s}},
\end{align*}

Next, on the set $\cbr{\omega\in \Omega: \theta_0(\omega) < +\ns}$ we define
\[
\vartheta_{1}:=\inf\{t\ge0 : |A|_t - |A|_{\theta_0}  \geq q\},
\
\sigma_{1}:=\inf\left\{ t\ge0 :\max_{j=1,2,\ldots,d} \rbr{\sbr{X^j}_{t} - \sbr{X^j}_{\theta_0}} \ge r \right\},
\]
otherwise we define $\theta_1 = +\ns$. Next we set 
\[
\theta_{1}:=\vartheta_{1} \wedge \sigma_{1},
\]
and introduce the following operator $T^{1}$:,
\begin{align*}
\left(T^{1}\mathbf G \right)_{t}: & =\mathbf Y_{t\wedge\theta_{0}}^0+\int_{t\wedge\theta_{0}}^{t\wedge\theta_{1}}K\left(s,\mathbf G\right)\dd \mathbf A_{s}+\int_{t\wedge\theta_{0}}^{t\wedge\theta_{1}}F\left(s,\mathbf G\right)\dd{\mathbf X_{s}}.
\end{align*}
Similarly as before, we prove that $T^{1}:{{\cal G}}\ra{{\cal G}}$, $T^{1}$ is a contraction and has
a fixed point $\mathbf Y^{1}\in {{\mathcal G}}$. Moreover, $\mathbf Y^{0}$ and $\mathbf Y^{1}$
agree on the interval $\sbr{0, \theta_0} \setminus\cbr{+\ns}$ and thus $\mathbf Y^{1}$ for any $t \in [0,  +\ns) $ satisfies 
\begin{align*}
{\mathbf Y}^1_{t \wedge \theta_1} & ={\mathbf Y}_{0}+\int_{0}^{t\wedge \theta_1}K\left(s, {\mathbf Y^1}\right)\dd {\mathbf A}_{s}+\int_{0}^{t\wedge \theta_1}F\left(s, {\mathbf Y^1}\right)\dd{{\mathbf X}_{s}}.
\end{align*}
Similarly, having defined  $\theta_{n}$, $T^{n}:{{\cal G}\ra{{\cal G}}}$,
and its fixed point $\mathbf Y^{n}$, $n=0,1,\ldots$ we define  the stopping time $\theta_{n+1}$ 
and introduce the operator $T^{n+1}:{{\cal G}}\ra{{\cal G}}$,
\begin{align*}
\left(T^{n+1}\mathbf G \right)_{t}: & =\mathbf Y_{t\wedge\theta_{n}}^{n}+\int_{t\wedge\theta_{n}}^{t\wedge\theta_{n+1}}K\left(s,\mathbf G\right)\dd \mathbf A_{s}+\int_{t\wedge\theta_{n}}^{t\wedge\theta_{n+1}}F\left(s,\mathbf G\right)\dd{\mathbf X_{s}}
\end{align*}
and its fixed point $\mathbf Y^{n+1}$,  which agrees with $\mathbf Y^{n}$ on the interval
$
\left[0,\theta_{n} \right] \setminus\cbr{+\ns}.
$

Finally, setting
\[
\mathbf Y:=\lim_{n\ra+\ns} \mathbf Y^{n}
\]
we get that for any $t \in [0,  +\ns) $ and $n\in \N$, $\mathbf Y$ satisfies
\begin{align*}
{\mathbf Y}_{t \wedge \theta_n} & ={\mathbf Y}_{0}+\int_{0}^{t\wedge \theta_n}K\left(s, {\mathbf Y}\right)\dd {\mathbf A}_{s}+\int_{0}^{t\wedge \theta_n}F\left(s, {\mathbf Y}\right)\dd{{\mathbf X}_{s}}.
\end{align*}

Now we will prove that $\lim_{n \ra +\ns} \theta_n(\omega) = +\ns$ for all $\omega \in \Omega$. Let us notice that for any $T>0$ from the inequality $\theta_{n}(\omega) \le T$, $n \in \N$, it follows that $|A|_T  (\omega) + \sbr{X}_T (\omega) \ge \min\rbr{q,r}\cdot n$. Since $|A|$ and $[X]$ are continuous for all $\omega \in \Omega$ (we choose such version of $[X]$), thus $|A|_T (\omega)$ and $\sbr{X}_T (\omega)$ are finite for all $\omega \in \Omega$ and
\begin{align*}
& \cbr{\omega \in \Omega: \lim_{n \ra +\ns} \theta_n(\omega) < +\ns} = \bigcup_{N=1}^{+\ns}\cbr{\omega \in \Omega: \lim_{n \ra +\ns} \theta_{n}(\omega) \le N} \\
& \subseteq \bigcup_{N=1}^{+\ns} \rbr{ \bigcap_{n=1}^{+\ns} \cbr{\omega \in \Omega: |A|_N  (\omega) + \sbr{X}_N (\omega) \ge \min\rbr{q,r} n } } = \emptyset.
\end{align*}

\subsubsection{Uniqueness}

In general, we can not guarantee that $\mathbf Y \in{{\cal G}}$, because we do not control the growth of the process $\mathbf A$. However, we have just proved that it is a solution in the sense of Definition \ref{solution_def}. Now we will prove that any two such solutions must be equal w.i.e. 

Let $\mathbf G$ and $\mathbf H$ be two solutions of \eqref{eq:equation} satisfying, together with sequences of stopping times $\rbr{\gamma_n }$ and $\rbr{\eta_n}$ respectively, conditions of Definition \ref{solution_def}. Let us define  $\tilde{\theta}_0 = \theta_0 \wedge \gamma_0 \wedge \eta_0 $ and \begin{align*}
\left(\tilde{T}^{0}{\mathbf G}\right)_{t} & ={\mathbf Y}_{0}+\int_{0}^{t\wedge\tilde{\theta}_{0}}K\left(s, {\mathbf G}\right)\dd {\mathbf A}_{s}+\int_{0}^{t\wedge\tilde{\theta}_{0}}F\left(s, {\mathbf G}\right)\dd{{\mathbf X}_{s}}. \end{align*}
Similarly as in \eqref{eq:Lipschitz} we prove that 
\begin{equation} \label{indish}
\overline{\E}\left(\tilde{T}^{0}{\mathbf G}-\tilde{T}^{0}{\mathbf H}\right)^{*}_{\cdot \wedge \tilde{\theta}_{0}} \le \frac{2}{3} \overline{\E}\left({\mathbf G}-{\mathbf H}\right)^{*}_{\cdot \wedge \tilde{\theta}_{0}}.
\end{equation}
On the other hand, since $\mathbf G$ and $\mathbf H$ are solutions of \eqref{eq:equation}, $\overline{\E}\rbr{ \tilde{T}^{0}{\mathbf G}_{\cdot \wedge \tilde{\theta}_{0}} - {\mathbf G}_{\cdot \wedge \tilde{\theta}_{0}} }^* = 0$, $\overline{\E}\rbr{ \tilde{T}^{0}{\mathbf H}_{\cdot \wedge \tilde{\theta}_{0}} - {\mathbf H}_{\cdot \wedge \tilde{\theta}_{0}} }^* = 0$. From this and \eqref{indish} we get $\overline{\E}\rbr{ {\mathbf G}_{\cdot \wedge \tilde{\theta}_{0}} - {\mathbf H}_{\cdot \wedge \tilde{\theta}_{0}} }^* = 0$.

Next, defining  $\tilde{\theta}_1 = \theta_1 \wedge \gamma_1 \wedge \eta_1 $,  
\begin{align*}
\left(\tilde{T}^{1}{\mathbf G}\right)_{t} & ={\mathbf G}_{t \wedge \tilde{\theta}_0}+\int_{t\wedge\tilde{\theta}_{0}}^{t\wedge\tilde{\theta}_{1}}K\left(s, {\mathbf G}\right)\dd {\mathbf A}_{s}+\int_{t\wedge\tilde{\theta}_{0}}^{t\wedge\tilde{\theta}_{1}}F\left(s, {\mathbf G}\right)\dd{{\mathbf X}_{s}}. \end{align*}
and reasoning similarly as before we get $\overline{\E}\rbr{ {\mathbf G}_{\cdot \wedge \tilde{\theta}_{1}} - {\mathbf H}_{\cdot \wedge \tilde{\theta}_{1}} }^* = 0$. 

Similarly, for $\tilde{\theta}_n = \theta_n \wedge \gamma_n \wedge \eta_n$ we get $\overline{\E}\rbr{ {\mathbf G}_{\cdot \wedge \tilde{\theta}_{n}} - {\mathbf H}_{\cdot \wedge \tilde{\theta}_{n}} }^* = 0$. Since $\tilde{\theta}_n(\omega) \ra +\ns$ as $n \ra +\ns$ for all $\omega \in \Omega$, we have 
\[
 \overline{\E}\rbr{ {\mathbf G} - {\mathbf H}}^* \le \sum_{n=1}^{+\ns} \overline{\E}\rbr{ {\mathbf G}_{\cdot \wedge \tilde{\theta}_{n}} - {\mathbf H}_{\cdot \wedge \tilde{\theta}_{n}} }^* = 0.
\]

\paragraph*{Acknowledgments.}

The work of Lesiba Ch. Galane and Farai J. Mhlanga was supported
in part by the National Research Foundation of South Africa (Grant
Number: 105924). The work of Rafa{\l } M. {\L ochowski} was
partially funded by the National Science Centre, Poland, under Grant
No.~$2016/21/$B/ST$1/01489$ and Grant
No.~$2019/35/$B/ST$1/042$. Part of this work was done while R.
M. {\L }. was visiting the Univeristy of Limpopo. The warm hospitality
of the Univeristy of Limpopo is gratefully acknowledged. The authors are grateful to Vladimir Vovk and \foreignlanguage{polish}{Adam Os"ekowski} for valuable questions and comments.

 \bibliographystyle{plain}
\bibliography{/Users/rafallochowski/biblio/biblio}

\end{document}